\documentclass[12pt]{article}

\usepackage{amssymb,amsmath,amscd,amsthm}

\newtheorem{theorem}{Theorem}[section]

\newtheorem{lemma}[theorem]{Lemma}
\newtheorem{proposition}[theorem]{Proposition}
\newtheorem{definition}[theorem]{Definition}

\begin{document}

\title{The Existence of Pair Potential Corresponding to Specified Density and Pair Correlation}

\author{ Leonid Koralov\footnote{Partially supported by NSF Research Grant} \\[1pt]
\normalsize Department of Mathematics\\[-4pt] \normalsize
Princeton University\\[-4pt] \normalsize Princeton, NJ
08544\\[-4pt] \normalsize koralov@math.princeton.edu\\[-4pt] }

\date{}
\maketitle

\begin{abstract}
Given a potential of pair interaction and a value of activity, one
can consider the Gibbs distribution in a finite domain $\Lambda
\subset \mathbb{Z}^d$. It is well known that for small values of
activity there exist the infinite volume ($\Lambda \rightarrow
\mathbb{Z}^d$) limiting Gibbs distribution and the infinite volume
correlation functions. In this paper we consider the converse
problem - we show that given $\rho_1$ and $\rho_2(x)$, where
$\rho_1$ is a constant and $\rho_2(x)$ is a function on
$\mathbb{Z}^d$, which are sufficiently small, there exist a pair
potential and a value of activity, for which $\rho_1$ is the
density and $\rho_2(x)$ is the pair correlation function.
\end{abstract}
\medskip
 {\bf Key words}: Gibbs Field, Gibbs Measure, Cluster Functions,
Pair Potential, Corrlation Functions, Ursell Functions.

\medskip
{\bf MSC Classification}: 60G55, 60G60.

\date{Received 3 December, 2004}

\section {Introduction}
\label{se1} Let us consider a translation invariant measure $\mu$
on the space of particle configurations on the lattice $
\mathbb{Z}^d$. For a given configuration each site can be occupied
by one particle or be empty. An $m$-point correlation function
$\rho_m(x_1,...,x_m)$ is the probability of finding $m$ different
particles at positions $x_1,...,x_m \in \mathbb{Z}^d$. The
following natural question has been extensively discussed in
physical and mathematical literature: given $\rho_1(x_1) \equiv
\overline{\rho}_1$ and $\rho_2(x_1,x_2) = \overline{\rho}_2(x_1 -
x_2)$, does there exist a measure $\mu$, for which these are the
first correlation function (density) and the pair correlation
function, respectively?

In the series of papers \cite{L1}-\cite{L3} Lenard provided a set
of relations on the functions $\rho_m$ which are necessary and
sufficient for the existence of such a measure. However, given
$\rho_1$ and $\rho_2$, it is not clear how to check if there are
some $\rho_3, \rho_4, ...$ for which these relations hold.

There are several recent papers which demonstrate the existence of
particular types of point processes (measures on the space of
particle configurations), which correspond to given $\rho_1$ and
$\rho_2$ under certain conditions on $\rho_1$ and $\rho_2$. In
particular, one dimensional point processes of renewal type are
considered  by Costin and Lebowitz in \cite{CL}, while
determinantal processes are considered by Soshnikov in \cite{So}.
In \cite{AS} Ambartzumian and Sukiasian prove the existence of a
point process corresponding to a sufficiently small density and
correlation function. Recently Costin and Lebowitz suggested
generalizations of their results. In \cite{ST} Stillinger and
Torquato consider fields over a space with finitely many points.
Besides, for the lattice model, they discuss possible existence of
a pair potential for a given density and correlation function
using cluster expansion without addressing the issue of
convergence.

In this paper we show that if $\rho_1$ and $\rho_2$ are small (in
a certain sense), there exists a measure on the space of
configurations for which $\rho_1$ is the density and $\rho_2$ is
the pair correlation function. Moreover, this measure is the Gibbs
measure corresponding to some pair potential and some value of
activity. In a sense, this is the converse of the classical
statement that a given potential of pair interaction and a
sufficiently small value of activity determine a translation
invariant Gibbs measure on the space of particle configurations in
$ \mathbb{Z}^d$ (or $ \mathbb{R}^d$) and the sequence of infinite
volume correlation functions.

\section{Notations and Formulation of the Result}
\label{se2}

We shall consider the following lattice system. Let $ \Phi(x)$, $x
\in \mathbb{Z}^d$ be a potential of pair interaction and let
$U(x_1,...,x_n) = \sum_{1 \leq i < j \leq n} \Phi(x_i - x_j)$ be
the total potential energy of the configuration $(x_1,...,x_n)$.
We assume that $\Phi(x) = \Phi(-x) \geq c_0 > -\infty$ for all $x$
and that $\Phi(0) = +\infty$.
 The full list of assumptions on $\Phi(x)$ will be given
below.

Let $ \Lambda $ be a finite subset of $ \mathbb{Z}^d$. The grand
canonical ensemble is defined by a measure on $ \bigcup_{n =
0}^\infty \Lambda^n$, whose restriction on $\Lambda^n$ is equal to
\[
\nu (x_1,...,x_n) = \frac{z^n}{n !} e^{-U(x_1,...,x_n)}~.
\]
The parameter $z >0$ is called the activity. The inverse
temperature, which is the factor usually present in front of the
function $U$, is set to be equal to one (or, equivalently,
incorporated into the function $U$). The total mass of the measure
is the grand partition function
\[
\Xi(\Lambda, z, \Phi) = \sum_{n = 0}^\infty \frac{z^n}{n !}
\sum_{(x_1,...,x_n) \in \Lambda^n}  e^{-U(x_1,...,x_n)}~.
\]
The $m$-point correlation function is defined as the probability
of finding $m$ different particles at positions $x_1,...,x_m \in
\Lambda$,
\[
\rho^{\Lambda}_m(x_1,...,x_m)  = \Xi(\Lambda, z, \Phi)^{-1}
\sum_{n = 0}^\infty \frac{z^{m+n}}{n !} \sum_{(y_1,...,y_n) \in
\Lambda^n} e^{-U(x_1,...,x_m,y_1,...,y_n)}~.
\]
The corresponding measure on the space of all configurations of
particles on the set $\Lambda$ (Gibbs measure) will be denoted by
$\mu^\Lambda$. Given another set $\Lambda_0 \subseteq \Lambda$, we
can consider the measure $\mu^{\Lambda}_{\Lambda_0}$ obtained as a
restriction of the measure $\mu^\Lambda$ to the set of particle
configurations on $\Lambda_0$.

 Given a potential of pair interaction $\Phi(x)$, we define $g(x)
= e^{-\Phi(x)} - 1$, $x \in \mathbb{Z}^d$. We shall make the
following standard assumptions:
\begin{equation} \label{assum1}
g(x) \geq -a > -1~~~ {\rm for}~~ x \neq 0.
\end{equation}
\begin{equation}
\label{assum2}
 g(0) = -1;~~~~g(x) = g(-x) ~~~ {\rm for}~~{\rm all}~~x;~~~\sum_{x \neq 0} |g(x)| \leq c < \infty.
\end{equation}
Clearly, any function $g(x)$ which satisfies
(\ref{assum1})-(\ref{assum2}) defines a potential of pair
interaction via
\[
\Phi(x) = - \ln(g(x) +1)~.
\]

It is well known (\cite{A}, \cite{B}) that when $\Lambda
\rightarrow \mathbb{Z}^d$ in a suitable manner (for example,
$\Lambda = [-k, k]^d$ and $k \rightarrow \infty$) the following
two limits exist  for sufficiently small $z$: \\ (a) There is a
probability  measure $\mu^{\mathbb{Z}^d}$ on the space of all
configurations on $\mathbb{Z}^d$, such that
\begin{equation} \label{limit1}
\mu^\Lambda_{\Lambda_0} \rightarrow
\mu^{\mathbb{Z}^d}_{\Lambda_0}~~~{\rm as}~~ \Lambda \rightarrow
\mathbb{Z}^d
\end{equation}
for any finite set $\Lambda_0 \subset \mathbb{Z}^d$.\\
 (b) All the correlation functions converge to
the infinite volume correlation functions. Namely,
\begin{equation} \label{limit2}
\rho^{\Lambda}_m(x_1,...,x_m) \rightarrow
\rho_m(x_1,...,x_m)~~~{\rm as} ~~ \Lambda \rightarrow
\mathbb{Z}^d.
\end{equation}
The infinite volume correlation functions are the probabilities
with respect to the measure $\mu^{\mathbb{Z}^d}$ of finding $m$
different particles at positions $x_1,...,x_m \in \mathbb{Z}^d$.
To make these statements precise we formulate them as a lemma.
\begin{lemma} (\cite{A}, \cite{B})
Assuming that (\ref{assum1}) and (\ref{assum2}) hold, there is a
positive $ \overline{z} = \overline{z}(a,c)$, such that
(\ref{limit1}) and (\ref{limit2}) hold for all $0 < z \leq
 \overline{z}$ when
$\Lambda = [-k, k]^d$ and $k \rightarrow \infty$.
\end{lemma}
Thus,   a pair potential defines a sequence of infinite volume
correlation functions for sufficiently small values of activity.
Note that $\rho_m(x_1,...,x_m) = 0$ if $x_i = x_j$ for $i \neq j$,
since two distinct particles can not occupy the same position.
Also note that all the correlation functions are translation
invariant,
\[
\rho_m(x_1,...,x_m) = \rho_m(0, x_2 - x_1,...,x_m -x_1)~.
\]
Thus, $\rho_1$ is a constant, $\rho_2$ can be considered as a
function of one variable, etc. Let $\overline{\rho}_m$ be the
function of $m-1$ variables, such that
\begin{equation} \label{sme}
\rho_m(x_1,...,x_m) = \overline{\rho}_m (x_2 - x_1,...,x_m -x_1)~.
\end{equation}

The main result of this paper is the following theorem.
\begin{theorem} \label{main}
Let $0 <  r < 1$ be a constant. Given any sufficiently small
constant $\overline{\rho}_1$ and any function
$\overline{\rho}_2(x)$, such that $\overline{\rho}_2(0) = 0$ and
$\sum_{x \neq 0} |\overline{\rho}_2(x) - \overline{\rho}_1^2| \leq
r \overline{\rho}_1^2$, there are a potential $\Phi(x)$, which
satisfies (\ref{assum1})-(\ref{assum2}), and a value of activity
$z$, such that $\overline{\rho}_1$ and $\overline{\rho}_2(x)$ are
the first and the second correlation functions respectively for
the system defined by $(z, \Phi)$.
\end{theorem}
\noindent {\bf Remark 1.} Let $\xi(x)$ be a random field with
values $0$ and $1$ (which is the same as a measure on the space of
particle configurations), and let $\overline{\rho}_1$ and
$\overline{\rho}_2(x)$ be its first two correlation functions.
Then
\[
E(\xi(x)- \overline{\rho}_1)(\xi(0) - \overline{\rho}_1) = \left\{
\begin{array}{ll}
                 \overline{\rho}_1 - \overline{\rho}_1^2 ~~~~~~~ \mbox{if $x = 0$}\\
                 \overline{\rho}_2(x) - \overline{\rho}_1^2 ~~~ \mbox{otherwise.}
            \end{array}
            \right.
\]
The positive definiteness of this function, which is necessary for
the existence of the field $\xi(x)$ with the given
$\overline{\rho}_1$ and $\overline{\rho}_2(x)$, is clearly
guaranteed by the conditions of the theorem if $\overline{\rho}_1$
is sufficiently small.

\noindent {\bf Remark 2.} As will seen from the proof of the
theorem, the pair potential and the activity corresponding to
given $\overline{\rho}_1$ and $\overline{\rho}_2(x)$ are unique,
if we restrict consideration to sufficiently small values of
$\Phi$ and $z$. The method of the proof allows one to explore the
properties of the pair potential based on the properties of the
correlation function.
\\

 The outline of the proof is the following. In Sections
\ref{clust} and \ref{erp}, assuming that a pair potential and a
value of the activity exist, we express the correlation functions
(or, rather, the cluster functions, which are closely related to
the correlation functions) in terms of the pair potential and the
activity. This relationship can be viewed as an equation for
unknown $\Phi$ and $z$. In Section \ref{mainr} we use the
contracting mapping principle to demonstrate that this equation
has a solution. In Section \ref{urs} we provide the technical
estimates needed to prove that the right hand side of the equation
on $\Phi$ and $z$ is indeed a contraction.

\section{Cluster Functions and  Ursell Functions}
\label{clust}

In this section we shall obtain a useful expression for cluster
functions in terms of the pair potential. The cluster functions
are closely related to the correlation functions.
 Some of the general known facts will be
stated in this section without proofs. The reader is referred to
Chapter 4 of \cite{A} for a more detailed exposition.

Let $A$ be the complex vector space of sequences $\psi$,
\[
\psi = (\psi_m(x_1,...,x_m))_{m \geq 0}
\]
such that, for each $m \geq 1$, $\psi_m$ is a bounded function on
$\mathbb{Z}^{md}$, and $\psi_0$ is a complex number. It will be
convenient to represent a finite sequence $(x_1,...,x_m)$ by a
single letter $X = (x_1,...,x_m)$. We shall write
\[
\psi(X)  = \psi_m(x_1,...,x_m).
\]
Let now $\psi^1, \psi^2 \in A$. We define
\[
\psi^1 \ast \psi^2 (X) = \sum_{Y \subseteq X} \psi^1(Y) \psi^2(X
\backslash Y),
\]
where the summation is over all subsequences $Y$ of $X$ and $X
\backslash Y$ is the subsequence of $X$ obtained by striking out
the elements of $Y$ in $X$.

Let $A_+$ be the subspace of $A$ formed by the elements $\psi$
such that $\psi_0 = 0$. Let $ \mathbf{1}$ be the unit element of
$A$ ( $ \mathbf{1}_0 = 1, \mathbf{1}_m \equiv 0$ for $m \geq 1$).

We define the mapping $\Gamma$ of $A_+$ onto $\mathbf{1} + A_+$ :
\[
\Gamma \varphi = \mathbf{1} + \varphi + \frac{\varphi \ast
\varphi}{2!} + \frac{\varphi \ast \varphi \ast \varphi}{3!} + ...
\]
The mapping $\Gamma$ has an inverse $\Gamma^{-1}$ on $\mathbf{1} +
A_+$:
\[
\Gamma^{-1}(\mathbf{1} + \varphi') = \varphi' - \frac{\varphi'
\ast \varphi'}{2} + \frac{\varphi' \ast \varphi' \ast \varphi'}{3}
- ...
\]
It is easy to see that $\Gamma \varphi (X)$ is the sum of the
products $\varphi(X_1)...\varphi(X_r)$ corresponding to all the
partitions of $X$ into subsequences $X_1,...,X_r$. If $\varphi \in
A_+$ and $\psi = \Gamma \varphi$, the first few components of
$\psi$ are
\[
\psi_0 = 1;~~~~\psi_1(x_1) = \varphi_1(x_1);~~~~\psi_2(x_1, x_2) =
\varphi_2(x_1, x_2) + \varphi_1(x_1) \varphi_1(x_2).
\]
Let $\Phi$ be a pair correlation function which satisfies
(\ref{assum1})-(\ref{assum2}), and let $z \leq \overline{z}(a,c)$.
Note that the sequence of correlation functions $\rho =
({\rho_m})_{m \geq 0}$ (with $\rho_0 = 1$) is an element of
$\mathbf{1} + A_+$.
\begin{definition}
The cluster functions $\omega_m(x_1,...,x_m)$, $m \geq 1$ are
defined by
\[
\omega = \Gamma^{-1} \rho.
\]
\end{definition}
Thus,
\[
\omega_1(x_1) = \rho_1(x_1);~~~~\omega_2(x_1,x_2) =
\rho_2(x_1,x_2) - \rho_1(x_1) \rho_1(x_2),
\]
or, equivalently,
\[
\overline{\omega}_1 = \overline{\rho}_1;~~~~\overline{\omega}_2(x)
= \overline{\rho}_2(x) - \overline{\rho}_1^2~
\]
where $\overline{\omega}_m$ are defined as in (\ref{sme}).

Let $\psi \in \mathbf{1} + A_+$ be defined by
\[
\psi_0 = 1;~~~~\psi_m(x_1,...,x_m) = e^{-U(x_1,...,x_m)}~.
\]
Define also
\[
\varphi = \Gamma^{-1} \psi~.
\]
\begin{definition}
The functions $\psi_m$ and $\varphi_m$ are called Boltzmann
factors and Ursell functions, respectively.
\end{definition}
\begin{lemma}
\label{expansion} (\cite{A}) The cluster functions can be
expressed in terms of the Ursell functions as follows
\[
\omega_m(x_1,...,x_m) = z^m \sum_{n = 0}^{\infty} \frac{z^n}{n!}
\sum_{y_1,...,y_n \in \mathbb{Z}^{d}}
\varphi_{m+n}(x_1,...,x_m,y_1,...,y_n)~.
\]
\end{lemma}
We shall later need certain estimates on the Ursell functions in
terms of the potential. To this end we obtain a recurrence formula
on a set of functions related to the Ursell functions. Given $X =
(x_1,...,x_m)$, we define the operator $D_X: A \rightarrow A$ by
\[
(D_X \psi)_n (y_1,...,y_n) = \psi_{m+n}(x_1,...,x_m,y_1,...,y_n)~.
\]
Then define
\[
\widetilde{\varphi}_X = \psi^{-1} \ast D_X \psi~,
\]
where $\psi$ is the sequence of Boltzmann factors, and $\psi^{-1}$
is such that $\psi^{-1} \ast \psi = \mathbf{1}$. It can be seen
that
\begin{equation}
\label{rel1} \varphi_{1+n}(x_1,y_1,...,y_n) =
\widetilde{\varphi}_{x_1}(y_1,...,y_n)
\end{equation}
and that the functions $ \widetilde{\varphi}_X $ satisfy a certain
recurrence relation, which we state here as a lemma.
\begin{lemma} (\cite{A}) \label{recurr}
The functions $ \widetilde{\varphi}_X $  satisfy the following
recurrence relation
\begin{equation} \label{rec}
\widetilde{\varphi}_X (Y) = \exp(- \sum_{i = 2}^m \Phi(x_i- x_1))
\sum_{S \subseteq Y} \prod_{j, y_j \in S}( \exp(-\Phi(y_j -x_1))
-1) \widetilde{\varphi}_{S \cup X \backslash x_1} (Y \backslash
S)~,
\end{equation}
where $X  = (x_1,...,x_m)$, $m \geq 1$, $Y = (y_1,...,y_n)$, $n
\geq 0$, and $\widetilde{\varphi}_X (Y) = \mathbf{1}$ if $m = 0$.
\end{lemma}

\section{Equations Relating the Potential, the Activity, and the Cluster Functions}
\label{erp} In this section we shall recast the main theorem in
terms of the cluster functions and examine a system of equations,
which relates the first two cluster functions with the pair
potential and the activity.

First, Theorem \ref{main} can clearly be re-formulated as follows
\begin{proposition} \label{main2}
Let $0 < r < 1$ be a constant. Given any sufficiently small
constant $\overline{\omega}_1$ and any function
$\overline{\omega}_2(x)$, such that $\overline{\omega}_2(0) =
-\overline{\omega}_1^2$ and $\sum_{x \neq 0}
|\overline{\omega}_2(x)| \leq r \overline{\omega}_1^2$, there are
a potential $\Phi(x)$,  which satisfies
(\ref{assum1})-(\ref{assum2}), and a value of activity $z$, such
that $\overline{\omega}_1$ and $\overline{\omega}_2(x)$ are the
first and the second cluster functions respectively for the system
defined by $(z, \Phi)$.
\end{proposition}

Consider the power expansions for $\omega_1$ and $\omega_2$, which
are provided by Lemma \ref{expansion}. Let us single out the first
term in both expansions. Note the translation invariance of the
functions $\omega_m$ and $\varphi_m$ and the fact that
$\varphi(x_1, x_2) = g(x_1 - x_2)$.
\begin{equation} \label{clust1}
\overline{\omega}_1 = z + z^2 \sum_{n=1}^\infty \frac{z^{n-1}}{n!}
\sum_{y_1,...,y_n \in \mathbb{Z}^{d}}
\varphi_{1+n}(0,y_1,...,y_n)~,
\end{equation}
\begin{equation} \label{clust2}
\overline{\omega}_2(x) = z^2 g(x) + z^3 \sum_{n=1}^\infty
\frac{z^{n-1}}{n!} \sum_{y_1,...,y_n \in \mathbb{Z}^{d}}
\varphi_{2+n}(0,x,y_1,...,y_n)~.
\end{equation}
Let
\[
A(z,g) = \sum_{n=1}^\infty \frac{z^{n-1}}{n!} \sum_{y_1,...,y_n
\in \mathbb{Z}^{d}} \varphi_{1+n}(0,y_1,...,y_n)~,
\]
\[
B(z,g)(x) = \sum_{n=1}^\infty \frac{z^{n-1}}{n!} \sum_{y_1,...,y_n
\in \mathbb{Z}^{d}} \varphi_{2+n}(0,x,y_1,...,y_n)~.
\]
Thus the equations (\ref{clust1}) and (\ref{clust2}) can be
rewritten as follows
\begin{equation} \label{eqn1}
z = \overline{\omega}_1 - z^2 A(z,g)~,
\end{equation}
\begin{equation} \label{eqn2}
g = \frac{\overline{\omega}_2}{z^2} - z B(z,g)~.
\end{equation}
Instead of looking at (\ref{eqn1})-(\ref{eqn2}) as a formula
defining $\overline{\omega}_1$ and $\overline{\omega}_2$ by a
given pair potential and the activity, we can instead consider the
functions $\overline{\omega}_1$ and $\overline{\omega}_2$ fixed,
and $g$ and $z$ unknown. Thus, Proposition \ref{main2} follows
from the following.
\begin{proposition} \label{main3}
If $\overline{\omega}_1$ and $\overline{\omega}_2$ satisfy the
assumptions of Proposition \ref{main2}, then the system
(\ref{eqn1})-(\ref{eqn2}) has a solution $(z,g)$, such that the
function $g$ satisfies (\ref{assum1})-(\ref{assum2}) and $z \leq
\overline{z}(a,c)$.
\end{proposition}
\section{Proof of the Main Result}
\label{mainr} This section is devoted to the proof of Proposition
\ref{main3}. We shall need the following notations. Let $
\mathcal{G}$ be the space of functions $g$, which satisfy
(\ref{assum2}) with some $c < \infty$. Let $||g|| = \sum_{x \neq
0} |g(x)|$. This is not a norm, since $ \mathcal{G}$ is not a
linear space, however ${d(g_1,g_2) = ||g_1-g_2||}$ is a metric on
the space $ \mathcal{G}$. Let $ \mathcal{G}_c$ be the set of
elements of $ \mathcal{G}$ for which $||g|| \leq c$. Note that if
$c <1$ then all elements of $ \mathcal{G}_c$ satisfy
(\ref{assum1}) with $a =c$.

We also define $I^{a_1, a_2}_{z_0} = [a_1 z_0, a_2 z_0]$. Let $D =
I^{a_1, a_2}_{z_0} \times  \mathcal{G}_c$. Note that if $c < 1$
then $(z,g) \in D$ implies that $z \leq \overline{z}(c,c)$ if
$z_0$ is sufficiently small. Thus, the infinite volume correlation
functions and cluster functions are correctly defined for $(z,g)
\in D$ if $z_0$ is sufficiently small.

 Let us define
an operator $Q$ on the space of pairs $(z,g) \in D$ by $Q(z,g) =
(z', g')$, where
\begin{equation} \label{eqn1a}
z' = \overline{\omega}_1 - z^2 A(z,g)~,
\end{equation}
\begin{equation} \label{eqn2a}
g'(x) = \frac{\overline{\omega}_2(x)}{z^2} - z B(z,g)(x)~~~~{\rm
for}~~x \neq 0;~~~~g'(0)= -1~.
\end{equation}
We shall prove the following lemma.
\begin{lemma} \label{l1}
Let $0 < r < 1$ be a constant. There exist positive constants $a_1
< 1$, $a_2 >1$, and $c < 1$ such that the equation $(z,g) =
Q(z,g)$ has a solution $(z,g) \in D$  for all sufficiently small
$z_0$ if $\overline{\omega}_1 = z_0$, $\overline{\omega}_2(0) =
-z_0^2$, and $\sum_{x \neq 0} |\overline{\omega}_2(x)| \leq r
z_0^2$.
\end{lemma}

Before we prove this lemma, let us verify that it implies
Proposition \ref{main3}. Let $0 < r < 1$ be fixed and let
$\overline{\omega}_1$ be sufficiently small for the statement of
Lemma \ref{l1} to be valid. Let $\overline{\omega}_2$ be such that
$\overline{\omega}_2(0) = -\overline{\omega}_1^2$ and $\sum_{x
\neq 0} |\overline{\omega}_2(x)| \leq r \overline{\omega}_1^2$.
Let $(z,g)$ be the solution of $(z,g) = Q(z,g)$, whose existence
is guaranteed by Lemma \ref{l1}. Let $\overline{\omega}'_1$ and
$\overline{\omega}'_2$ be the first two cluster functions
corresponding to the pair $(z,g)$. Note that $\overline{\omega}_1$
and $\overline{\omega}'_1$ satisfy the same equation
\[
z = \overline{\omega}_1 - z^2 A(z,g)~;~~~~
 z = \overline{\omega}'_1
- z^2 A(z,g)~.
\]
Therefore, $\overline{\omega}_1 = \overline{\omega}'_1$. The
functions  $\overline{\omega}_2$ and $\overline{\omega}'_2$ also
satisfy the same equation
\[
g(x) = \frac{\overline{\omega}_2(x)}{z^2} - z B(z,g)(x)~;~~~~ g(x)
= \frac{\overline{\omega}'_2(x)}{z^2} - z B(z,g)(x)~;~~~~ {\rm
for}~~x \neq 0.
\]
Thus, $\overline{\omega}_2(x) = \overline{\omega}'_2(x)$ for $x
\neq 0$. The fact that $\overline{\omega}_2(0) =
\overline{\omega}'_2(0)$ follows from
\[
\overline{\omega}_2(0) = - \overline{\omega}_1^2 =   -
{\overline{\omega}'_1}^2 = \overline{\omega}'_2(0)~.
\]
Thus it remains to prove Lemma \ref{l1}. The proof will be based
on the fact that for small $z_0$ the operator $Q: D \rightarrow D$
is a contraction in an appropriate metric.  Define
\[
d_{z_0}(z_1,z_2) = \frac{h |z_1 -z_2|}{z_0}~.
\]
The value of the constant $h$ will be specified later. Now the
metric on $D$ is given by
\[
\rho((z_1,g_1),(z_2,g_2)) =  d_{z_0}(z_1,z_2) + d(g_1,g_2)~.
\]
Lemma \ref{l1} clearly follows from the contracting mapping
principle and the following lemma
\begin{lemma} \label{l2}
Let $0 < r < 1$ be a constant. There exist positive constants $a_1
< 1$, $a_2 >1$, and $c < 1$ such that for all sufficiently small
$z_0$ the operator $Q$ acts from the domain $D$ into itself and is
uniformly contracting in the metric $\rho$ for some value of
$h>0$, provided that $\overline{\omega}_1 = z_0$,
$\overline{\omega}_2(0) = -z_0^2$, and $\sum_{x \neq 0}
|\overline{\omega}_2(x)| \leq r z_0^2$.
\end{lemma}
\proof Take $c = \frac{r +2}{3}$, $a_1 =\sqrt{\frac{2r}{r+1}}$,
$a_2 = 2$.
We shall need certain estimates on the values of $A(z,g)$ and
$B(z,g)$ for $(z,g) \in D$. Namely, there exist universal
constants $u_1,...,u_6$, such that for sufficiently small $z_0$ we
have
\begin{equation} \label{co1}
\sup_{(z,g) \in D}|A(z,g)| \leq u_1~.
\end{equation}
\begin{equation} \label{co2}
\sup_{(z,g) \in D}\sum_{x \neq 0}|B(z,g)(x)| \leq u_2~.
\end{equation}
\begin{equation} \label{co3}
\sup_{(z_1,g), (z_2,g) \in D}|A(z_1,g)- A(z_2,g)| \leq u_3 |z_1 -
z_2| ~.
\end{equation}
\begin{equation} \label{co4}
\sup_{(z,g_1), (z,g_2) \in D}|A(z,g_1)- A(z,g_2)| \leq u_4
d(g_1,g_2)~.
\end{equation}
\begin{equation} \label{co5}
\sup_{(z_1,g), (z_2,g) \in D}\sum_{x \neq 0}|B(z_1,g)(x)-
B(z_2,g)(x)| \leq u_5 |z_1 - z_2|~.
\end{equation}
\begin{equation} \label{co6}
\sup_{(z,g_1), (z,g_2) \in D}\sum_{x \neq 0}|B(z,g_1)(x)-
B(z,g_2)(x)| \leq u_6 d(g_1,g_2)~.
\end{equation}
These estimates follow from Lemma \ref{estu} below. For now,
assuming that they are true, we continue with the proof of Lemma
\ref{l2}. The fact that $Q D \subseteq D$ is guaranteed by the
inequalities
\begin{equation} \label{in1a}
z_0 + (a_2 z_0)^2 u_1 \leq a_2 z_0~,
\end{equation}
\begin{equation} \label{in2a}
z_0 - (a_2 z_0)^2 u_1 \geq a_1 z_0~,
\end{equation}
\begin{equation} \label{in3a}
\frac{r z_0^2}{(a_1 z_0)^2} + a_2 z_0 u_2 \leq c~.
\end{equation}
It is clear that (\ref{in1a})-(\ref{in3a}) hold for sufficiently
small $z_0$. Let us now demonstrate that  for some $h$ and for all
sufficiently small $z_0$ we have
\begin{equation} \label{contr}
\rho( Q(z_1,g_1), Q(z_2, g_2)) \leq \frac{1}{2} \rho( (z_1,g_1),
(z_2, g_2))~~~{\rm if}~~~  (z_1,g_1), (z_2,g_2) \in D.
\end{equation}
  First, taking (\ref{co1}), (\ref{co3}), and (\ref{co4}) into
account, we note that
\[
d_{z_0}(z_1^2 A(z_1,g_1), z_2^2 A(z_2,g_2)) \leq d_{z_0}(z_1^2
A(z_1,g_1), z_2^2 A(z_1,g_1)) +
\]
\[
d_{z_0}(z_2^2 A(z_1,g_1), z_2^2 A(z_2,g_1))+  d_{z_0}(z_2^2
A(z_2,g_1), z_2^2 A(z_2,g_2)) \leq
\]
\[
\frac{u_1 h |z_1^2 - z_2^2|}{z_0} + \frac{u_3 h (a_2 z_0)^2 |z_1 -
z_2|}{z_0} + \frac{u_4 h (a_2 z_0)^2 d(g_1,g_2)}{z_0}~.
\]
If  $h$ is fixed, the right hand side of this inequality can be
estimated from above, for all sufficiently small $z_0$, by
\[
\frac{1}{6}(d_{z_0}(z_1,z_2) + d(g_1,g_2)).
\]
Similarly,
\[
\sum_{x \neq 0}|z_1 B(z_1,g_1)(x) - z_2 B(z_2,g_2)(x)| \leq
\sum_{x \neq 0}|z_1 B(z_1,g_1)(x) - z_2 B(z_1,g_1)(x)| +
\]
\[
\sum_{x \neq 0}|z_2 B(z_1,g_1)(x) - z_2 B(z_2,g_1)(x)| + \sum_{x
\neq 0}|z_2 B(z_2,g_1)(x) - z_2 B(z_2,g_2)(x)| \leq
\]
\[
u_2|z_1 - z_2| + u_5 a_2 z_0 |z_1 - z_2| + u_6 a_2 z_0
d(g_1,g_2)~.
\]
Again, if $h$ is fixed, the right hand side of this inequality can
be estimated from above, for all sufficiently small $z_0$, by
\[
\frac{1}{6}(d_{z_0}(z_1,z_2) + d(g_1,g_2)).
\]
Finally,
\[
\sum_{x \neq 0}|\frac{\overline{\omega}_2(x)}{z_1^2} -
\frac{\overline{\omega}_2(x)}{z_2^2} | \leq r z_0^2
|\frac{1}{z_1^2} - \frac{1}{z_2^2}| \leq \frac{2 a_2 |z_1 -
z_2|}{a_1^4 z_0}~.
\]
We can now take $h = \frac{12 a_2}{a_1^4}$, which implies that the
right hand side of the last inequality can be estimated from above
by $\frac{1}{6} d_{z_0}(z_1,z_2)$. We have thus demonstrated the
validity of (\ref{contr}), which means that the operator $Q$ is
uniformly contracting. This completes the proof of the lemma. \qed
\section{Estimates on the Ursell Functions} \label{urs}
In this section we shall derive certain estimates on the Ursell
functions, which, in particular, will imply the inequalities
(\ref{co1})-(\ref{co6}).
\begin{lemma} \label{estu}
Suppose that the functions $g_1(x)$ and $ g_2(x)$ satisfy
(\ref{assum2}) with $c < 1$. Let
$\varphi^k=(\varphi^k_m(x_1,...,x_m))_{m \geq 0}$, $k = 1,2$ be
the corresponding Ursell functions. Then there exist constants
$q_1$ and $q_2$ such that
\[
\sum_{y_1,...,y_n \in \mathbb{Z}^{d}}
|\varphi^k_{1+n}(0,y_1,...,y_n)| \leq n! q_1^{n+1}~,~~~k = 1,2,
\]
\[
\sum_{y_1,...,y_n \in \mathbb{Z}^{d}}
|\varphi^1_{1+n}(0,y_1,...,y_n)- \varphi^2_{1+n}(0,y_1,...,y_n) |
\leq n! q_2^{n+1} ||g_1 -g_2||~.
\]
\end{lemma}
Note that the inequalities (\ref{co1})-(\ref{co6}) immediately
follow from this lemma and the definitions of $A(z,g)$ and
$B(z,g)(x)$.

Recall that in Section \ref{clust} we introduced the functions
$\widetilde{\varphi}_X(Y)$, which were closely related to the
Ursell functions. Given $g_1(x)$ and $ g_2(x)$ which satisfy
(\ref{assum2}) with $c < 1$, we now define
\[
r^k(m,n) = \sup_{(x_1,...,x_m)} \sum_{y_1,...,y_n \in
\mathbb{Z}^{d}}
|\widetilde{\varphi}^k_{(x_1,...,x_m)}(y_1,...,y_n)|~,~~~k=1,2,
\]
\[
d(m,n) = \sup_{(x_1,...,x_m)} \sum_{y_1,...,y_n \in
\mathbb{Z}^{d}}
|\widetilde{\varphi}^1_{(x_1,...,x_m)}(y_1,...,y_n)-
\widetilde{\varphi}^2_{(x_1,...,x_m)}(y_1,...,y_n)|~.
\]
We shall prove the following lemma.
\begin{lemma} \label{estu2}
Suppose that the functions $g_1(x)$ and $ g_2(x)$ satisfy
(\ref{assum2}) with $c < 1$.  Then there exist constants $q_1$ and
$q_2$ such that
\begin{equation} \label{er}
r^k(m,n) \leq n! q_1^{m+n}~,~~~k = 1,2,
\end{equation}
\begin{equation} \label{eq}
d(m,n) \leq n! q_2^{m+n} ||g_1 -g_2||~.
\end{equation}
\end{lemma}
Sine we can express the Ursell functions in terms of
$\widetilde{\varphi}_X(Y)$ via (\ref{rel1}), Lemma \ref{estu2}
immediately implies Lemma \ref{estu}. It remains to prove Lemma
\ref{estu2}.\\ \\ {\it Proof of Lemma \ref{estu2}.} The estimate
(\ref{er}) follows from (4.27) of \cite{A}, and thus we shall not
prove it here. We proceed with the proof of (\ref{eq}).

 In the definition of $d(m,n)$ we can take the
supremum over a restricted set of sequences $(x_1,...x_m)$, namely
those sequences, for which all $x_i$ are distinct. Indeed, if $x_i
= x_j$ for $i \neq j$, then
$\widetilde{\varphi}^1_{(x_1,...,x_m)}(y_1,...,y_n) =
\widetilde{\varphi}^2_{(x_1,...,x_m)}(y_1,...,y_n) = 0$, as
follows from the definition of $\widetilde{\varphi}_X(Y)$.

Let $f_k(x) =  e^{-\Phi_k(x)} = g_k (x) +1~$, $k =1,2$. We shall
need the fact that if $\mathcal{X}$ is any set, which does not
contain $x_1$, then
\[
\prod_{x \in \mathcal{X}} f_k(x -x_1) \leq \exp({  \sum_{x \in
\mathcal{X}} \ln (g_k(x -x_1) +1) }) \leq \exp(  \sum_{x \in
\mathcal{X}}g_k(x -x_1) )\leq e^c~.
\]
The proof of (\ref{eq}) will proceed via an induction on $m+n$.
Assume that $x_1,...,x_m$ are all distinct. From the recurrence
relation (\ref{rec}) it follows that
\[
 \sum_{y_1,...,y_n \in
\mathbb{Z}^{d}}
|\widetilde{\varphi}^1_{(x_1,...,x_m)}(y_1,...,y_n)-
\widetilde{\varphi}^2_{(x_1,...,x_m)}(y_1,...,y_n)| =
\]
\[
 \sum_{y_1,...,y_n \in
\mathbb{Z}^{d}} | \prod_{i = 2}^m f_1(x_i- x_1) \sum_{S \subseteq
Y} \prod_{j, y_j \in S} g_1(y_j -x_1) \widetilde{\varphi}^1_{S
\cup X \backslash x_1} (Y \backslash S) -
\]
\[
 \prod_{i = 2}^m f_2(x_i- x_1) \sum_{S \subseteq Y} \prod_{j,
y_j \in S} g_2(y_j -x_1) \widetilde{\varphi}^2_{S \cup X
\backslash x_1} (Y \backslash S)| \leq I_1 + I_2,
\]
where
\[
I_1 =  \sum_{y_1,...,y_n \in \mathbb{Z}^{d}}  \sum_{S \subseteq Y}
| \prod_{i = 2}^m f_1(x_i- x_1) \prod_{j, y_j \in S} g_1(y_j -x_1)
( \widetilde{\varphi}^1_{S \cup X \backslash x_1} (Y \backslash S)
- \widetilde{\varphi}^2_{S \cup X \backslash x_1} (Y \backslash
S))|~,
\]
\[
I_2 = \sum_{y_1,...,y_n \in \mathbb{Z}^{d}}  \sum_{S \subseteq Y}
|[ \prod_{i = 2}^m f_1(x_i- x_1)  \prod_{j, y_j \in S} g_1(y_j
-x_1) -
\]
\[
  \prod_{i = 2}^m
f_2(x_i- x_1)  \prod_{j, y_j \in S} g_2(y_j -x_1)]
\widetilde{\varphi}^2_{S \cup X \backslash x_1} (Y \backslash
S)|~.
\]
Note that there are $\frac{n!}{s!(n-s)!}$ subsequences $S$ of the
sequence $Y$, which are of length $s$. Rearranging the sum, so
that to take it over all possible values of $s$, we see that
\[
I_1 \leq \sum_{s = 0}^n  \frac{n!}{s!(n-s)!} \sum_{y_1,...,y_s \in
\mathbb{Z}^{d}}
 | \prod_{i = 2}^m f_1(x_i- x_1) \prod_{j =
1}^s g_1(y_j -x_1)| d(m+s -1, n-s) \leq
\]
\[
 \sum_{s = 0}^n  \frac{n!}{s!(n-s)!} e^c(1+ c)^s  d(m+s -1, n-s)~.
\]
Similarly,
\[
I_2 \leq \sum_{s = 0}^n  \frac{n!}{s!(n-s)!} \sum_{y_1,...,y_s \in
\mathbb{Z}^{d}}  | \prod_{i = 2}^m f_1(x_i- x_1) \prod_{j = 1}^s
g_1(y_j -x_1) -
\]
\[
\prod_{i = 2}^m f_2(x_i- x_1) \prod_{j = 1}^s g_2(y_j -x_1) |
r(m+s-1, n-s)~.
\]
Let \[ F_k^{a,b} = \prod_{i = a}^b |f_k (x_i- x_1)|,~~~~{\rm
where}~~~2 \leq a \leq b \leq m~~~{\rm and}~~~k  = 1,2, \]
\[ G_k^{a,b} = \prod_{i = a}^b |g_k (y_i- x_1)|,~~~~{\rm
where}~~~1 \leq a \leq b \leq s~~~{\rm and}~~~k  = 1,2. \] Note
that
\begin{equation} \label{fest}F_k^{a,b} \leq e^c~,
\end{equation}
\begin{equation}
\label{gest}  \sum_{y_a,...,y_b \in \mathbb{Z}^{d}}  G_k^{a,b}
\leq (1+c)^{b-a+1}~.
\end{equation}

Then,
\[
 \sum_{y_1,...,y_s \in
\mathbb{Z}^{d}}  | \prod_{i = 2}^m f_1(x_i- x_1) \prod_{j = 1}^s
g_1(y_j -x_1) - \prod_{i = 2}^m f_2(x_i- x_1) \prod_{j = 1}^s
g_2(y_j -x_1) | \leq
\]
\[
 \sum_{y_1,...,y_s \in
\mathbb{Z}^{d}} [ |f_1(x_2-x_1) - f_2(x_2 -x_1)|F_1^{3,m}G_1^{1,s}
+
\] \[F_2^{2,2}|f_1(x_3-x_1) - f_2(x_3-x_1)| F_1^{4,m} G_1^{1,s} +
...+F_2^{2,m-1}|f_1(x_m - x_1) - f_2(x_m - x_1)| G_1^{1,s}+
\]
\[
F_2^{2,m}|g_1(y_1 - x_1) - g_2(y_1 - x_1)| G_1^{2,s} + ...+
F_2^{2,m} G_2^{1,s-1} |g_1(y_s - x_1) - g_2(y_s - x_1)|].
\]
There are $m+s$ terms inside the square brackets. In addition to
(\ref{fest}) and (\ref{gest}) we use the fact that
\[
|f_1(x_i - x_1) - f_2(x_i - x_1)| \leq ||g_1 - g_2||~,~~~~2 \leq i
\leq m, \] \[
 \sum_{y_i \in \mathbb{Z}^{d}} |g_1(y_i - x_1) -
g_2(y_i - x_1)| \leq ||g_1 - g_2||~,~~~~1 \leq i \leq n.
\]
Therefore, the entire sum can be estimated from above by
\[
(m+s) e^{2c} (1 +c)^s ||g_1 - g_2||.
\]
Therefore,
\[
I_2 \leq \sum_{s = 0}^n  \frac{n!}{s!(n-s)!}  (m+s) e^{2c} (1
+c)^s ||g_1 - g_2|| r(m+s-1, n-s) \leq
\]
\[
||g_1 - g_2|| (m+n) e^{2c} n! q_1^{m+n -1}  \sum_{s = 0}^n
\frac{(1+c)^s}{s!} \leq ||g_1 - g_2|| (m+n) e^{1+3c} n! q_1^{m+n
-1}~.
\]
Combining this with the estimate on $I_1$ we see that
\[
d(m,n) \leq  \sum_{s = 0}^n  \frac{n!}{s!(n-s)!} e^c(1+ c)^s d(m+s
-1, n-s) +
\]
\[
||g_1 - g_2|| (m+n) e^{1+3c} n! q_1^{m+n -1}~.
\]
Let us use induction on $m+n$ to prove that
\begin{equation}
\label{ind}  d(m,n) \leq n! q_2^{m+n} ||g_1 -g_2|| (m+n)
\end{equation}
for some value of $q_2$. The statement is obviously true for $m+n
= 0$. Assuming that the induction hypothesis holds for all $m',n'$
with $m'+n' \leq m+n -1$, we obtain
\[
d(m,n) \leq  \sum_{s = 0}^n  \frac{n!}{s!(n-s)!} e^c(1+ c)^s
(n-s)! q_2^{m+n-1} ||g_1 -g_2||(m+n-1) +
\]
\[
||g_1 - g_2|| (m+n) e^{1+3c} n! q_1^{m+n -1} \leq
\]
\[
||g_1 - g_2|| (m+n) e^{1+3c} n!( q_1^{m+n -1} + q_2^{m+n -1}).
\]
The expression in the right hand side of this inequality is
estimated from above by the right hand side of (\ref{ind}) if $q_2
= 2 e^{1+3c} \max(1,q_1)$. Thus, (\ref{ind}) holds for all $m,n$
with this choice of $q_2$. Note that we can get rid of the factor
$(m+n)$ in the right hand side of (\ref{ind}) by taking a larger
value of $q_2$. This completes the proof of (\ref{eq}) and of
Lemma \ref{estu2}. \qed \vspace{1cm}
 {\bf Acknowledgements } I would like to
express my gratitude to J. Lebowitz, S. Torquato, and particularly
Y. Sinai for a number of very useful discussions.


\begin{thebibliography}{9999}

\bibitem{AS} R. Ambartzumian, H. Sukiasian. {\it
Inclusion-Exclusion and Point Processes}. Acta Applicandae
Mathematicae 22; pp 15-31, 1991.
\bibitem{CL} O. Costin, J. Lebowitz. {\it On the Construction of
Particle Distributions with Specified Single and Pair Densities.}
Journal of Physical Chemistry B. 108 (51) (2004), 19614-19618
(cond-mat0405519).
\bibitem{L1} A. Lenard. {\it Correlation Functions and the
Uniqueness of the State in Classical Statistical Mechanics.}
Commun. Math. Phys. 30, pp 35-44 (1973).
\bibitem{L2} A. Lenard. {\it States of Classical Statistical Mechanical
Systems of Infinitely Many Particles. I.} Arch. Rational Mech.
Anal. 59 (1975), no. 3, 219--239.
\bibitem{L3} A. Lenard. {\it States of Classical Statistical Mechanical
Systems of Infinitely Many Particles. II.} Arch. Rational Mech.
Anal. 59 (1975), no. 3, 241--256.
\bibitem{B} R. Minlos. {\it Limiting Gibbs Distribution.}
Funktsional'nyi Analiz i Ego Prilozheniya, Vol. 1, No. 2, pp.
60-73, March-April 1967.
\bibitem{A} D. Ruelle. {\it Statistical Mechanics. Rigorours
Results.} W.A. Benjamin, inc., 1969.
\bibitem{So} A. Soshnikov. {\it Determinantal Random Point Fields.}
Russian Mathematical Surveys, vol.55, No.5, pp.923-975, (2000)
\bibitem{ST} F. Stillinger, S. Torquato. {\it Pair Correlation
Function Realizability: Lattice Model Implications.} Journal of
Physical Chemistry B, 108 (51) (2004) 19589-19594.

\end{thebibliography}
\end{document}